\documentclass[12pt,onecolumn, a4paper]{article}
\usepackage {amsfonts}
\usepackage [english]{babel}
\usepackage [utf8]{inputenc}
\usepackage [T1]{fontenc}
\usepackage{setspace}
\usepackage {graphicx}
\usepackage {rotate}
\usepackage{authblk}
\usepackage{anysize}
\usepackage{mdwlist}
\usepackage[compact]{titlesec}
\usepackage{titling}
\usepackage[sc]{mathpazo}
\usepackage{lipsum}
\usepackage{multicol}

\posttitle{\par\end{center}\vspace{-1.6em}}

\titleformat{\chapter}{\centering\large}{\thechapter.}{2.5em}{}
\titleformat{\section}{\large\bfseries}{\thesection}{2em}{}
\titleformat{\subsection}{\normalsize\bfseries}{\thesubsection}{1.5em}{}
\titleformat{\subsubsection}{\normalsize\it}{\thesubsubsection}{1em}{}
\titlespacing{\section}{2pt}{*3}{*1}
\titlespacing{\subsection}{15pt}{*2}{*0}
\titlespacing{\subsubsection}{22pt}{*2}{*1}

\marginsize{2.5cm}{2.5cm}{2.5cm}{2.5cm}
\title{Agent-Based Stock Market Model with Endogenous Agents' Impact}
\author{Jan A. Lipski\thanks{corresponding author: j.lipski@student.uw.edu.pl} }
\author{Ryszard Kutner}
\affil{Faculty of Physics, University of Warsaw: \\ Hoża 69, PL-00681 Warsaw, Poland}

\renewcommand{\thefigure}{\arabic{figure}}

\begin{document}
\maketitle

\begin{abstract}
The three-state agent-based 2D model of financial markets as proposed by Giulia Iori has been extended by introducing an increased trust in the correctly predicting agents, a more realistic consultation procedure as well as a formal validation mechanism. This paper shows that such a model correctly reproduces the three fundamental stylised facts: fat-tail log returns, power-law volatility autocorrelation decay over time and volatility clustering.
\end{abstract}

\section{Introduction}
 The non-Brownian dynamics of financial markets has been observed in their modern study. Agent-based modelling is an attempt to mimic real markets, especially the fundamental stylised facts: fat-tail (non-Gaussian) log returns distribution, power-law absolute log returns (volatility) autocorrelation decay in time and time clustering of high log returns and volatility. In such models individual investors make up the financial market and determine the price of the stock they trade. 
 
This paper extends our agent-based model \cite{My}, which is based on an established model by Giulia Iori \cite{Iori}. Our initial improvement was to introduce endogenous herding behaviour by increasing the esteem of the correctly predicting agents and modelling other agents as their followers. The next improvement is to introduce a more realistic consultation procedure and to propose a formal validation mechanism for the fundamental stylised fact studied, that is, for the fat-tail distribution of log returns.
\section{Highlights of the Iori model}
The version of the three-state Ising model (also known as the Potts model) of financial markets, proposed by Iori \cite{Iori}, gave results reproducing the fundamental stylised facts: volatility clustering, fat-tail log-return distribution, and power-law decay in time of the volatility autocorrelation function. 

The system consists of $n$ agents placed in the nodes of of a 2D square lattice who determine the stock price $P(t)$ at time $t$. Every agent $i = 1,…,n$ can take one of three spin states representing three types of behaviours: $+1$ as buy, $0$ as stay inactive, $-1$ as sell and is under  the influence of a local field $Y_i$, being the sum of the forces exerted by its nearest neighbours $J_{ij}(t)$ and by a random noise $\eta_i(t)$:
\begin{equation}
\label{pole}
Y_i(t) = \sum\limits_{j = 1}^4 J_{ij}(t) + \eta_i(t). 
\end{equation}

The temporal spin of the $i$th agent, $\sigma_i$ is governed by the rule:

\begin{equation}
\label{spiny}
\sigma_i(t+1) = \left\{ \begin{array}{rl}
1 &\mbox{if $Y_i(t) \geq \xi_i(t),$}\\
0 &\mbox{if $-\xi_i(t) < Y_i(t) < \xi_i(t)$}\\
-1 &\mbox{if $Y_i(t) \leq -\xi_i(t)$,}
\end{array} \right. ,
\end{equation}
where $\xi_i$ is the individual threshold, which is initially drawn from a standard normal distribution and later thus adjusted:
\begin{equation} \label{2}
\xi_i(t) = \xi_i(t-1) \frac{P\left(t-1\right)}{P\left(t-2\right)}, \ t \geq 2.
\end{equation}

In the Iori model, the price $P(t)$ is determined as a function of supply $S(t)$ (number of negative spins) and demand $D(t)$ (number of positive spins) as follows:

\begin{equation}
\label{wzornacene}
P(t)= P\left(t-1\right) \left(\frac{D(t-1)}{S(t-1)}\right) ^{\kappa(t-1)},  
\end{equation}
where
\begin{equation}
\label{exp4}
  \kappa(t-1) = \alpha\frac{D(t-1)+S(t-1)}{n}.
\end{equation} 

It is worth noting that this definition is Iori's unique contribution, however the equation for price always mirrors the fundamental economic law of supply and demand as shown in \cite{Wlosi}, \cite{MDRK} (and refs. therein).

The stylised facts reproduced by our simulations refer to a log return $r$ defined over a time period $t-t'$ as a natural logarithm of the ratio of the corresponding prices $P(t)$ and $P(t-t')$:
\begin{equation}
\label{logret}
  r(t-t') = \ln\left(\frac{P(t)}{P(t-t')}\right),  \; t' \leq t.
\end{equation}

In this paper we further exploit the term volatility defined as the absolute value of the log return defined in the above Eq. (\ref{logret}). 

\section{Beyond the Iori model} \label{Beyond}
\subsection{Foreseeability of neighbours as a source of trust}
Our initial improvements, discussed in detail in \cite{My}, introduce a common sense relationship between the agents' decisions and their influence on their nearest neighbours as well as the presence of fundamental behaviour from time to time. 

The influence of an agent should be great if the agent foresees the price movement correctly, and should be small in the opposite case. That is, the influence increases if the product of an individual agent's spin value at a certain time in the past and the log return over that time is positive and decreases if that product is negative. Thus, the $J_{ij}$ force exerted by agent $j$ on agent $i$ (as in Eq. (\ref{pole})) is defined as:

\begin{equation}
\label{soins}
J_{ij}(t) = W_{ij}+\displaystyle\sum\limits_{\tau' = 
t - \tau}^{t - 1} \sigma_j(\tau') \ln\left(\frac{P(t - 1)}{P(\tau')}\right), \ \tau \geq 2.
\end{equation}

The coefficient $J_{ij}(t=0)$ is an initial value allotted at the beginning of the simulation: either 1 with a fixed probability $p$, or $0$ with a probability of $1-p$ (as in Iori's model) and $W_{ij}$ is the background static impact of agent $j$ on agent $i$, randomly drawn from a uniform unit distribution. The sum over $\tau'$ represents the altering component of the impact with the agents' memory being $\tau$ steps long.

\subsection{Semi-periodical fundamental behaviour}\label{semi-period}
To avoid a persistent positive feedback effect resulting in constantly directed price changes, the fundamental agents' behaviour is introduced. The fundamental behaviour is constrained by positive factors $a > 1$ and $b > 1$. If the market price is greater than the fundamental price by multiplier $a$, the agent sells shares. In the opposite case, if the market price is lower than the fundamental price by factor $b$, the agent buys shares. An individual agent behaves this way with a globally-defined probability $P_{fund}$. We assume that each agent knows the share fundamental price.


A detailed description of the algorithm which the simulation is based upon is described in our previous paper \cite{My}.

\section{Further improvements}

\subsection{Short consultation round}

The first improvement was a more realistic model of the consultation round. That is, the Iori model \cite{Iori} assumed that the agents can start trading only when stabilisation of their opinions occurs. However, in this paper we assumed that the agents trade directly after consulting their colleagues (the nearest neighbours). That is, they have time to consult only a few times (namely, make only a few phone calls) before trading. We assumed that no opinion stabilisation, formally referred to as system relaxation or thermalisation, needs to be achieved, as indeed no real stock market stops if investors are inactive.

For instance, we have proposed only four consultation rounds in which the agents can change their opinions. This improvement has led to more stable reconstruction of fat-tail log returns and time decay of absolute log returns autocorrelation.

\subsection{Validation}
A key problem in agent-based models is the models' validation and parameter estimation. In this section we present the tools for validation and estimation and in Section \ref{fattail} we present our results. 

As revised by Helbing \cite{Helbing}, a common approach is to check the stability of reconstruction of the fundamental stylised facts either by eye judgement or by using a formal statistical test. In this paper we propose to validate the log returns histogram, which, as commonly observed in stock markets, should exhibit a strong non-Gaussian behaviour. 

In this paper we use the Kolmogorov - Smirnov (K-S) test \cite{KSTest}, which is a further improvement of our initial model \cite{My}. The K-S  test 
is based upon computing the maximum distance between two (empirical or theoretical) cumulative distribution functions. In our case, the test compares two empirical CDFs of log returns. One of the CDFs describes the simulation results while the other one describes the historical log returns for Standard and Poor's 500 index in the period from January 2nd 1980 to May 10th 2013.

From the total number of 23 parameters, only two, the most significant, parameters have been chosen. The altering parameters were the probability $P_{fund}$ of the agents trading according to the stock fundamental value (cf. Section \ref{semi-period}) and the range of the system's memory $\tau$ (cf. Eq. (\ref{soins})).
\newpage
\section{Simulation results}
\subsection{Fat-tail log returns}\label{fattail}
The result of the reconstruction of fat-tail log returns distribution is quite stable. In fact, the results of the K-S test for the  altering parameters $P_{fund}$ (first column) and $\tau$ (second column) agree with the empirical data to a large extent, as shown in Tab. \ref{KSres}.

\begin{center}
\begin{table}[h]
\begin{center}
\begin{tabular}{|l|l|c|}
\hline
$P_{fund}$ & $\tau$ &   p-value \\ 
\hline
60 & 20  & 0.998881\\ 
60 & 24 & 0.999972 \\ 
60 & 32 & 0.986818 \\ 
70 & 20 & 0.999163 \\ 
70 & 24 & 0.999675 \\ 
70 & 28 & 0.999308 \\ 
70 & 32 & 0.999665 \\ 
80 & 20 & 0.999877 \\ 
80 & 24 & 0.999741 \\ 
80 & 36 & 0.999865 \\ 
80 & 40 & 0.999749 \\ 
90 & 20 & 0.999859 \\ 
90 & 24 & 0.999639 \\ 
90 & 40 & 0.988326 \\ 
\hline
\end{tabular}
\end{center}

\caption{P-values of the K-S test comparing the log returns c.d.f. of simulation results to the historical data for the simulations with different parameter values $P_{fund}$ and $\tau$.}\label{KSres}
\end{table}
\end{center}

As shown in Tab. \ref{KSres}, the agreement with empirical data is stable for different parameter values. The standard deviation of those p-values is small and equals $0.0044$. A typical plot, for parameters $P_{fund} = 90$ and $\tau = 24$, is shown in Fig. \ref{histograms}. As shown in  Fig. \ref{histograms}, the distribution of the historical data and simulation results exhibit a non-Gaussian behaviour. \newpage

 


\renewcommand{\thefigure}{\arabic{figure}}
\begin{figure}[h]\begin{center}
 \includegraphics[width=0.8\textwidth]{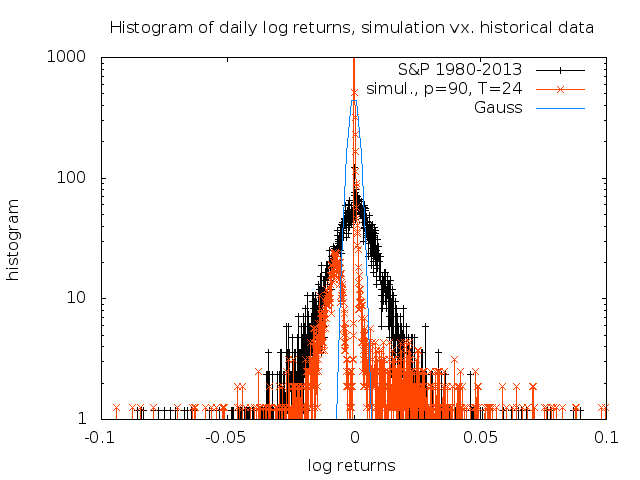}
\caption{Histograms of daily log returns obtained from historical data and from simulation for parameters $P_{fund} = 90$, $\tau = 24$.}\label{histograms}
\end{center}\end{figure}

%

\subsection{Power-law decay of volatility autocorrelation function} 
The (normalised) autocorrelation function (averaged over time $t$)  as a function of time lag $\tau$ is defined as follows:
\begin{equation}\label{problem}
 C(\tau) = \frac{\left<r(t) \, r(t-\tau)\right> - \left<r\right>^2}{Var(r)},  
\end{equation}
where $t$ is the decision round number, $\tau$ is the decision round time lag, $\left< ... \right>$ is the time average (cf. Eq. (\ref{logret})), and $Var(r)$ is the variance of the log return.

\renewcommand{\thefigure}{\arabic{figure}}
\begin{figure}[h]\begin{center}
 \includegraphics[width=0.8\textwidth]{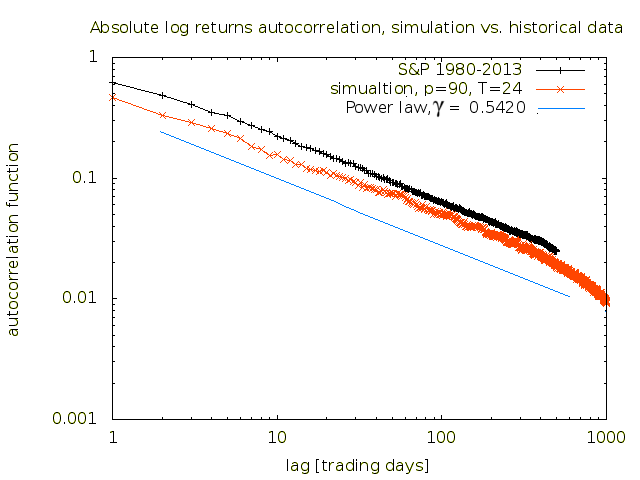}
\caption{Autocorrelation function decay of absolute log returns in time obtained from historical data and simulation results for parameters $P_{fund} = 90$, $\tau = 24$, compared with power law distribution (solid line).}\label{Autocorr}
\end{center}\end{figure}

The autocorrelation function of absolute daily log-returns reveals long-term power-law relaxation versus time:
\begin{equation}
\label{powlaw}
R(\tau) \propto \tau^{-\gamma}, \gamma > 0,
\end{equation}
where $R$ is the relaxation function of time lag $\tau$ and $\gamma$ is a positive exponent. This model gives astonishingly correct predictions for the values of $\gamma$ for the first $500$ time lag days (see Fig. \ref{Autocorr} for details). For the S\&P500 index $\gamma(SP500) = 0.5577$, for the simulation $\gamma(simul.) = 0.5420$, which gives off $97\%$ agreement. This is a successful confirmation of the validity of our model.

\subsection{Volatility clustering}
As shown below (see plots in Fig. \ref{cluster}) , we reproduced the temporal clustering of high and low log returns and high and low volatility quite well. Apparently, the log returns clustering obtained in the variogram of the simulation is not as pronounced as for empirical data. However, the regions of high and low absolute log returns (that is, high and low volatility) are clearly visible.
\renewcommand{\thefigure}{\arabic{figure}}
\begin{figure}[h]\begin{center}
 \includegraphics[width=0.8\textwidth]{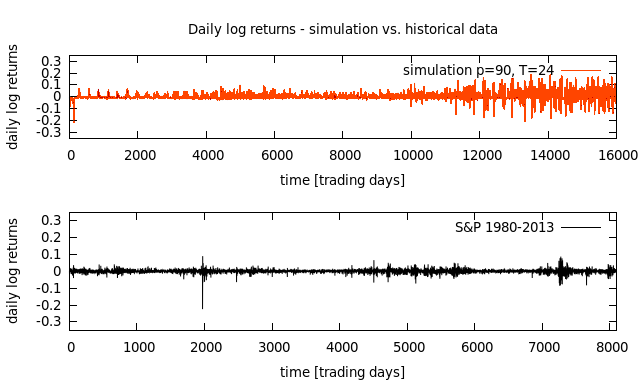}
\caption{Comparison of volatility clustering for historical data and results of the simulation with parameters $P_{fund} = 90$, $\tau = 24$}\label{cluster}
\end{center}\end{figure}
\newpage
\section{Conclusions}
In this paper we have proposed a sufficiently realistic and tractable agent-based model of financial markets. In our earlier paper \cite{My}, we extended the basic algorithm of the Iori model by introducing an increasing trust in agents who predict market movements correctly, and a decreasing trust in agents who fail to do so. However, in this paper we have proposed a more realistic scheme of consultation rounds between agents. The model has been extended to short and non-thermalising (non-equilibrium) consultation rounds. This has increased the validity of our highly non-Gaussian log return distributions.

We have also introduced a formal way of validating the model with the Kolmogorov-Smirnov test for different sets of parameters. It has also reproduced two other fundamental stylised facts: volatility and log returns clustering as well as the power-law decay of volatility autocorrelation. 

Apparently, we have achieved a stable reconstruction of the fundamental stylised facts corroborated with formal tests. Moreover, it is tractable and easy to alter model, which makes promising and capable predictions of real market price changes after further improvements.

\end{document}